# Manganese Dissolution in alkaline medium with and without concurrent oxygen evolution in LiMn$_2$O$_4$

Omeshwari Bisen[¶†], Max Baumung[¶†], Cynthia A. Volkert[†], Marcel Risch[¶†]*

† Institute of Materials Physics, University of Göttingen, Friedrich-Hund-Platz 1, 37077 Göttingen, Germany.

¶ Nachwuchsgruppe Gestaltung des Sauerstoffentwicklungsmechanismus, Helmholtz Zentrum Berlin für Materialien und Energie GmbH, Hahn-Meitner-Platz 1, 14109 Berlin, Germany.

## ABSTRACT

Manganese dissolution during the oxygen evolution reaction (OER) has been a persistent challenge that impedes the practical implementation of Mn-based electrocatalysts including the Li$_x$Mn$_2$O$_4$ system in aqueous alkaline electrolyte. The investigated LiMn$_2$O$_4$ particles exhibit two distinct Mn dissolution processes; one independent of OER and the other associated to OER. Combining the bulk sensitive X-ray absorption spectroscopy, surface sensitive X-ray photoelectron spectroscopy and electrochemical detection of Mn dissolution using rotating ring-disk electrode, we explore the less understood Mn dissolution mechanism during OER. We correlate near-surface oxidation with the charge attributed to dissolved Mn, which demonstrates increasing Mn dissolution with the formation of surface Mn$^{4+}$ species under anodic potential. The observed stronger dissolution during the OER is attributed to the formation of additional $Mn^{4+}$ from $Mn^{3+}$ during OER. We show that control over the amount of $Mn^{4+}$ in $Li_xMn_2O_4$ before the onset of the OER can partially mitigate the OER-triggered dissolution. Overall, our atomistic insights into the Mn dissolution processes are crucial for knowledge-guided mitigation of electrocatalyst degradation, which can be broadly extended to manganese-based oxide systems.

## INTRODUCTION

In the light of globally increasing energy consumption, stable and efficient electrocatalytic energy storage and conversion are paramount for the transition from dwindling fossil sources to sustainable sources owing to the intermittent nature of renewable energy.[1,2] Water splitting provides an attractive avenue for chemical energy storage (e.g. in hydrogen bonds) but it suffers from the inefficiency of the oxygen evolution reaction (OER).[3–6] In the last few decades, there has been tremendous attention on enhancing the activity of OER catalysts, but the stability of these catalysts has not been extensively investigated.[7] These challenges have persisted for an extended period, with limited attention directed towards comprehending the stability of electrocatalysts and elucidating the degradation mechanisms of the OER catalysts. These mechanisms encompass various processes such as catalyst dissolution, change in crystal phase and morphology, support passivation, detachment of the electrocatalyst from the electrode, and blocking of active sites.[7] Among these, dissolution of catalysts is generally believed to be one of the most important causes of catalysts degradation.[8] In most studies, the stability of catalysts is often inferred solely by change in electrolysis current density and/or overpotential values.[9–11] However, stable activity over time does not guarantee materials stability because of persistent dissolution of metal cations from an oxide surface, potentially causing dynamic surface changes and increased area as leached ions unveil new surfaces.[12]

The dissolution of Mn is a critical concern in Mn oxide-based electrocatalysis and battery research.[13–17] We note that Mn can be highly mobile on the surface of Mn oxides as recently reported by environmental TEM.[18] Four Mn dissolution processes have been discussed in literature previously: (1) the $MnO_2/MnO_4^-$ redox pair, (2) surface reconstruction, (3) bulk changes, and (4) the OER.[1,8,14,15,19,20] The $Mn^{4+}O_2/Mn^{7+}O_4^-$ redox is predicted by thermodynamics (commonly inferred from an E-pH/ Pourbaix diagram)[1,14,15]; its calculated reversible potential is +0.595 V vs SHE[21] ($MnO_4^- + 2H_2O + 3e^- \rightleftharpoons MnO_2(s) + 4OH^-$; 1.362 V vs. RHE at pH 13).[22,23] It has been experimentally observed at 1.41 V vs. RHE using ICP-MS.[14] Surface reconstruction cannot be predicted a priori. The relevant experimental potentials are between 0.67 and 1.525 V vs RHE where $Mn_3O_4$ with tetrahedral $Mn^{2+}$ is formed on Mn oxides as observed, e.g., by ICP-MS at 0.675 V vs. RHE.[15,24] Additionally, we have previously observed Mn loss due to bulk delithiation of $LiMn_2O_4$ in alkaline electrolytes as probed by RRDE.[20,24,25] Finally, metal dissolution may be coupled with the OER,[26] which also affects Mn oxides.[14,15] Since Mn oxidizes to $Mn^{4+}$ before the onset of the OER[24] and surface reconstruction is common during OER,[8] the Mn loss during OER may be related to the aforementioned processes that can also occur without OER. Experimentally, Mn loss during OER was observed by Rabe et al. in the range 1.525 to 1.75 V vs RHE.[15] Out of the four Mn dissolution processes, the one during OER is least understood, yet most relevant for OER electrocatalysis.

In this study, we use a pristine nano-sized cubic spinel $LiMn_2O_4$ particles as a model system for the OER in alkaline media to investigate the Mn dissolution processes during potential cycling. We used the Pt ring of a RRDE[25] to detect Mn dissolution current and explore it for mechanistic investigation of electrocatalyst degradation. Tafel slopes obtained from qualitative Mn dissolution current at the RRDE ring were used to investigate the two distinct Mn dissolution processes, one independent of OER and other associated to it. The Mn oxidation state and $LiMn_2O_4$ composition are elucidated using X-ray absorption spectroscopy and XPS, which we correlate with the mechanistic insights. Further, we elucidated the role of $Mn^{4+}$ species formed at the surface during anodic cycling with the Mn dissolution. Our results suggest electrocatalyst treatment prior to the OER as a strategy for mitigating electrocatalyst degradation by controlling the Mn oxidation state.

**RESULTS AND DISCUSSION**

We use a pristine $LiMn_2O_4$ powder as a model system for the OER in alkaline media to investigate the Mn dissolution processes during potential cycling. The tunable electronic properties of Mn in $LiMn_2O_4$ during OER cycling enable the insightful understanding of OER activity and stability of electrocatalysts. $LiMn_2O_4$ has a well-defined crystal structure (Figure S1, inset) with a cubic spinel lattice structure (Fd3m) which is confirmed by X-ray diffraction (XRD, Figure S1).[25,27]

For the electrochemical measurements, we have used a three-electrode cell with rotating ring-disk electrode (RRDE) in 0.1 M NaOH (pH=13), where $LiMn_2O_4$ decorated a glassy carbon (GC) rod as a disk electrode and an outer concentric Pt ring as a ring electrode. The Pt ring was used to qualitatively probe the Mn corrosion/dissolution current electrochemically by applying 1.2 V vs. RHE at Pt ring, which is thermodynamically favorable potential to reduce the $MnO_4^-$ or $MnO_4^{2-}$ moieties in electrolyte into some form of $MnO_x$[25] or by applying 0.4 V vs. RHE to qualitatively probe oxygen evolution by oxygen reduction. The detailed electrochemical procedure is documented in the experimental section. Figure 1(a) represents the CV of total disk current and as-measured ring current for oxygen and Mn detection at a scan rate of 10 mV/s for cycle 1. The first cycle exhibits the distinct onset of the total disk current, OER and Mn dissolution current. Notably, the disk current began to increase prior to 1.4 V vs RHE, whereas Mn ring current started to rise after 1.4 V vs RHE, which is well before the onset of oxygen evolution (above 1.6 V vs RHE). Thus, the disk current before 1.4 V vs RHE has solo contribution of the oxidation of Mn moieties, whereas after 1.4 V vs RHE, current due to Mn dissolution also arises. At voltage above 1.6 V vs RHE, we observed the exponential increase in OER ring current in addition to the Mn dissolution. This implies that the total disk current of the anodic scan arises from catalysis (i.e., $O_2$ detection), corrosion (Mn detection), and Mn oxidation current. The cathodic scans are exponential without any relevant feature. In the later cycle, the Mn dissolution current and Mn oxidation current drastically diminishes with cycling (Figure 1b), whereas the catalytic current alters little (Figure S2), which is in agreement with our previous work.[25]

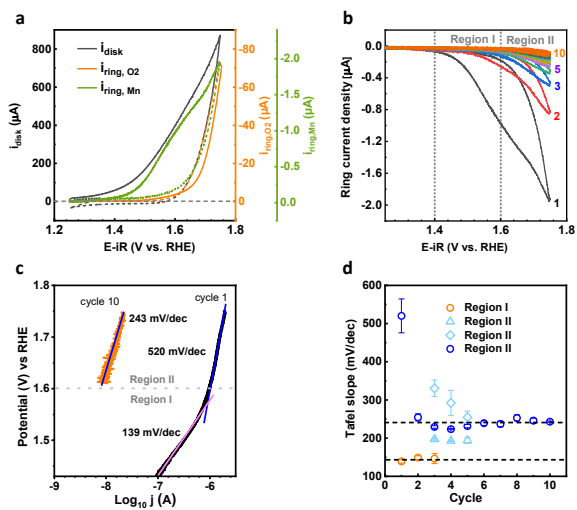

**Figure 1.** (a) Cyclic voltammetry of disk current and corresponding as-measured ring current of OER and Mn dissolution for cycle 1. The anodic scan is shown as solid lines and the cathodic scan as dashed lines. (b) As measured Mn corrosion current observed at Pt ring (potential applied at ring is 1.2 V vs RHE) for 1st to 10th cycle and (c) Tafel slope corresponding to 1st and 10th cycle. The two distinct Mn loss processes observed before and after OER onset (above1.6 V vs RHE) for 1st cycle. (d) Tafel slope as a function of number of Mn cycles.

With the aim of understanding the corrosion process for LiMn$_2$O$_4$, we have monitored the loss of Mn on the Pt ring of RRDE during 10 cycles (Figure 1b). For the first cycle, the CV with two onsets (E$_{Mn\ process1}$ ~ above 1.4 V and E$_{Mn\ process2}$ ~ above 1.6 V vs RHE) of exponentially increasing current was observed before and after the OER onset (above 1.6 V vs RHE). We have assigned the two distinct potential regions for investigating the Mn dissolution current, which represent the potential regime without (Region I) and with OER (Region II). The Mn dissolution processes corresponding to Region I and Region II are proposed as Process I and Process II, respectively. The CV of subsequent cycles looked similar but with lower current suggesting that Mn dissolution decreases relative to cycle 1 and in later cycles in both regions. While virtually no Mn dissolution current was detected in Region I during later cycles, it was still detected in Region II, which further supported distinct processes I and II.

The mechanistic insights from the distinct behaviors of Mn dissolution in Region I and Region II were better understood by Tafel analysis of the Mn dissolution current at the Pt ring (Figure 1c, Figure S3 and Table S1). The Tafel slope (b = $\partial$logi/$\partial$E) in chemical equilibrium indicates how the kinetic currents scale with the applied potential.[28] For cycle 1, we observed the two distinct slopes of 139±4 mV/dec in Region I and 520±44 mV/dec in Region II. The Tafel slope of 139±4 mV/dec in Region I is close to 120 mV/dec, which represents a reaction mechanism with electrochemical RLS (assuming a transfer coefficient of 0.5).[28–30] While Region II has the Tafel slope value of 520±44 mV/dec which is clearly higher value than 120 mV/dec. Such high values are commonly assigned to reaction mechanisms with chemical limiting steps without any pre-equilibrium reaction step involved prior to RLS.[28,29] Beyond the first cycle, the Tafel slope in Region I remained constant until it vanished after the 3$^{rd}$ cycle and the Tafel slope in Region II decreased to 240±18 mV/dec, which might indicate a change to an electrochemical limiting step with a small transfer coefficient (b=0.25).[29–31] Depending on the how many Tafel slopes are assumed, the decrease either happened gradually over 5 cycles or abruptly from the 1st to the 2nd cycle (Figure 1d). Nonetheless, we observe two distinct processes, namely Process I in Region I and Process II in Region II with different values and trends of the Tafel slopes. After potential cycling, only Process II remained active.

To understand the insights from the two distinct Mn dissolution processes, it is important to investigate both processes independently by experimental design where we investigate only Region I until a steady-state is reached (Figure 2a; "conditioning" experiment) and only then enter Region II (Figure 2c; "post conditioning" experiment). The anodic Mn ring current in Region I exhibited a Tafel slope of approximately 120 mV/dec for 10 cycles (details in Figure 2b and Table S2), which agreed with the data in a wider potential range shown in Figure 1b ("without conditioning" experiment). Subsequently enlarging the potential window to include Region II during post conditioning, we observed that the Tafel slope values around 200 mV/dec (Figure 2d and Table S3, note: two different Tafel slopes were observed for the early cycles), which agrees with the data in Figure 1b (without conditioning, after first cycle) where a wider potential range was used from the start. These experiments demonstrated that the processes can be separated and by comparison to Figure 1d that Process II in early cycles was coupled to Process I.

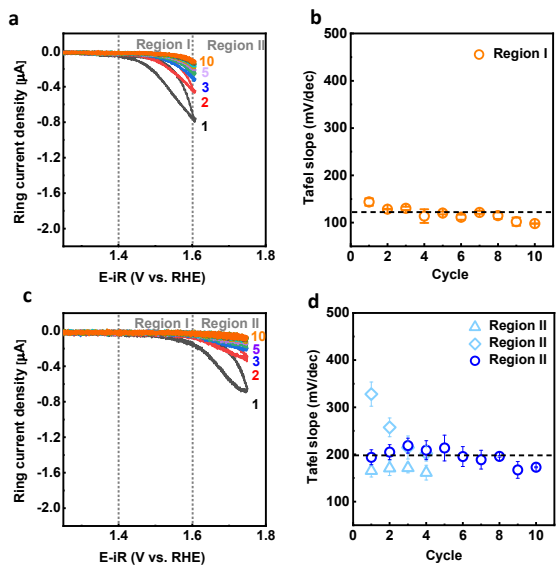

**Figure 2** (a) Mn corrosion current observed at Pt ring for 1st to 10th cycle before OER (in potential window of 1.25-1.6 V vs RHE) and (b) Tafel slope corresponding to it in region I. (c) Mn corrosion current observed at Pt ring for 1st to 10th cycle after the conditioning process (in potential window of 1.25-1.75 V vs RHE) and (d) corresponding Tafel slopes in region II.

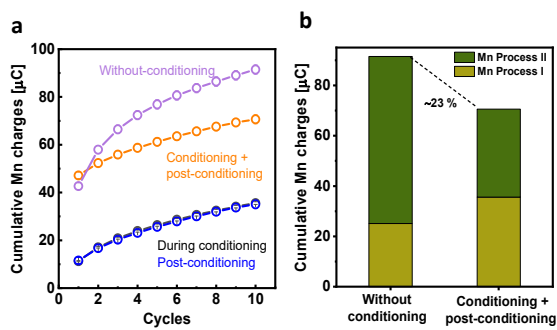

**Figure 3** (a) The cumulative Mn charges corresponding to without-conditioning, during conditioning and post-conditioning and conditioning + post-conditioning Mn dissolution current for 10 cycles. (b) Total Mn charges from 10 cycles corresponding to without-conditioning and conditioning + post-conditioning Mn dissolution current and different color corresponds to the distinct Mn dissolution process in region I and region II.

By using the insights from these two Mn processes, our aim was to mitigate and/or lower down the Mn loss in the OER regime. For evaluating progress toward this goal, we qualitatively investigated the Mn charge passed during the conditioning and post conditioning experiments (Figure 2) as well as compared it with the Mn charges during the experiment without conditioning (Figure 1b). We have measured the charge passed through Mn dissolution by integrating the current at the ring over time (Table S4). The cumulative charges during conditioning and post conditioning were identical within error and the cumulative charge of the sequence of these two experiments was lower than that of the experiment without conditioning after the $2^{nd}$ cycle (Figure 3a). This indicates that the conditioning step is beneficial to lower down the Mn loss from the $LiMn_2O_4$ system; the total Mn charges dissolved during 10 CVs with conditioning step is approximately 23% less in comparison to the CVs without conditioning (Figure 2b). The decrease in Mn loss is mainly due to the decrease in Mn dissolution in Region II (Process II) by ~ 47.29% (detailed analysis in SI, Table S4). No drastic variation in OER activity within the error (shown in Figure S4) and smaller Mn dissolution current after applying the conditioning, represents the early formation of more robust Mn states, which leads to the lowering the Mn dissolution. Furthermore, the observations from charge analysis suggests the interaction of distinct Mn dissolution processes during cycling, which led to higher dissolution compared to the case when treating them separately. This can be better understood by complementing electrochemical studies with the spectroscopic characterization techniques.

**Table 1.** Summary of the Samples ID and the description of its preparation procedure.

| Sample name | Preparation |
|---|---|
| **LMO** | As-received $LiMn_2O_4$ powder |
| **LMO_ink** | Dropcasted as LMO ink |
| **LMO_c0f_1.4V** | Dropcasted as LMO ink, then scan to 1.4 V at 10 mV/s |
| **LMO_c0f_1.6 V** | Dropcasted as LMO ink, then scan to 1.6 V at 10 mV/s |
| **LMO_c1f** | Dropcasted as LMO ink then single cycle from 1.25 V to 1.75 V and back to 1.25 V at 10 mV/s |
| **LMO_c10f or Without-conditioning** | Dropcasted as LMO ink then 10 cycles from 1.25 V to 1.75 V and back to 1.25 V at 10 mV/s |
| **LMO_c10c or conditioning** | Dropcasted as LMO ink then 10 cycles from 1.25 V to 1.6 V and back to 1.25 V at 10 mV/s |
| **LMO_c10c_c10f or post-conditioning** | Dropcasted as LMO ink, processed as LMO_c10c and then 10 cycles from 1.25 V to 1.75 V and back to 1.25 V at 10 mV/s |

**Understanding of structure-activity/stability relationships**

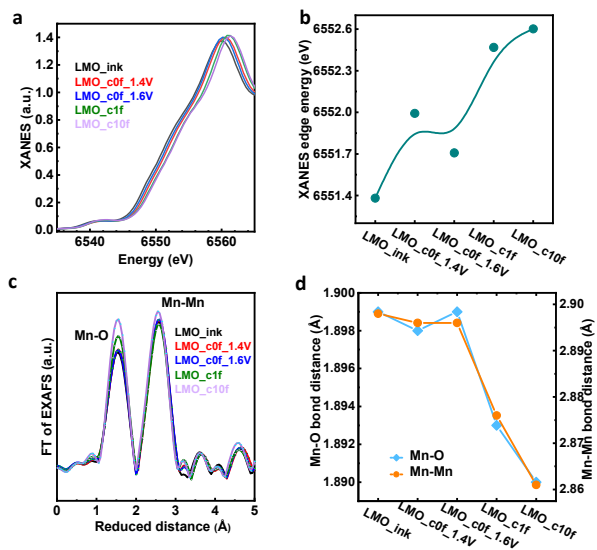

**FIGURES 4.** (a, b) XANES spectra at Mn-K edge and corresponding relation of XANES edge energy with LMO_ink and electrochemically processed samples, (c,d) FT of EXAFS spectra at the Mn-K edge and corresponding relation of Mn-O and Mn-O bond distances with LMO_ink and electrochemically processed samples.

Process I and Process II could produce different microstructures, in particular Mn oxidation states. We performed bulk sensitive X-ray absorption spectroscopy (XAS) and surface-sensitive X-ray photoelectron spectroscopy (XPS) after selected preparation condition of key points in the previously discussed experiments (Table 1 and Figure S5). Figure 4a shows Mn-K Edge X-ray absorption near-edge structure (XANES) spectra of aforementioned samples.[32] The edge position visually shifts to higher energy with higher potential and cycling as analyzed by the integral method[33] in Figure 4b. Sample LMO_c0f_1.6V is an exception to this trend, likely due to a slight variation during sample preparation, which does not affect our conclusions. Using a calibration curve assembled from Mn oxides with known oxidation states (Figure S6), we calculated that the formal oxidation state of Mn in sample LMO_ink is about +3.33(3), which is lower than a stoichiometric $LiMn_2O_4$ powder (+3.50) likely due to ink preparation. [34,35]The formal oxidation state increases from $Mn^{3.5+}$ to $Mn^{3.6+}$ in the order LMO_c0f_1.4V to LMO_c0f_1.6V to LMO_c1f to LMO_c10f (Table S5; note that the 1.4V and 1.6 V samples are identical within error).

The change in the local structure (mainly averaged interatomic distance) of the selected samples was further tracked by EXAFS (extended X-ray absorption fine structure) at the Mn-K edge.[32] The Fourier transform (FT) of the EXAFS of LMO_ink and electrochemically processed samples showed the two prominent peaks corresponding to Mn-O and Mn-Mn bonding of the spinel $LiMn_2O_4$ (Figure 4c) without additional peaks that would indicate new phases. In order to resolve the minor changes in Mn-O and Mn-Mn bond distances, we have carried out EXAFS fitting (Table S6 and Figure S7-S8). The trends of the Mn-O and Mn-Mn distances mirrored that of the edge position as expected. For higher potentials and cycles, both distances reduced (Figure 4d). The oxidation of Mn in $Li_xMn_2O_4$ is commonly coupled to delithiation (x<1), for which a reduction of the lattice parameter is expected.[36,37] Additionally, lattice contraction is expected for the loss of Mn detected in the electrochemical experiments (Figures 1 and 2).[37,38] This motivated us to check how the composition of the samples changed after key points of the electrochemical measurements.

XPS was used to track the Mn, Li and O content and the populations of Mn oxidation states in the near surface region of the samples after selected electrochemical treatment as defined in Table 1. Firstly, wide XPS spectra were recorded (Figure S9). They comprised the expected signal from Mn, Li and O and the additional signal from Na due to residuals of electrolyte after electrochemical processing. The presence of Na 1s signals makes the quantitative analysis of the Mn 2p, Mn 3p and Li 1s edges unreliable. Therefore, we have mainly focused on qualitative analysis of Mn 2p, Mn 3p and Li 1s via high resolution-XPS (HRXPS) by normalizing the Mn $2p_{3/2}$ peak and Mn 3p peak height. Figure S10a and S10b show the Mn $2p_{3/2}$ and Mn 3p/Li 1s HRXPS spectra of the samples without conditioning while Figures S10c and S10d show the respective HRXPS spectra of samples after conditioning and after post-conditioning. The sample LMO_ink exhibited a shoulder at the low binding energy (B.E.) side of the Mn $2p_{3/2}$ and Mn 3p peaks indicating the presence of some amount of $Mn^{2+}$, which is in well agreement with the EELS analysis and soft-XAS of the Mn-$L_3$ edge of pristine LMO where the tetrahedral antisite $Mn^{2+}$ defects are present in the near surface region.[36,39] All electrochemically treated samples have equal or lower height of the shoulder assigned to tetrahedral $Mn^{2+}$. Furthermore, the same trends are observed for the Li1s peak, which was highest for sample LMO_ink and decreased with electrochemical treatment, indicating delithiation. Further changes in the main peak position of Mn $2p_{3/2}$ and shape, which we ascribe to the population of $Mn^{3+}$ and $Mn^{4+}$ moieties, which is also in well agreement with the delithiation. All qualitative changes in the spectra are summarized in Table 2.

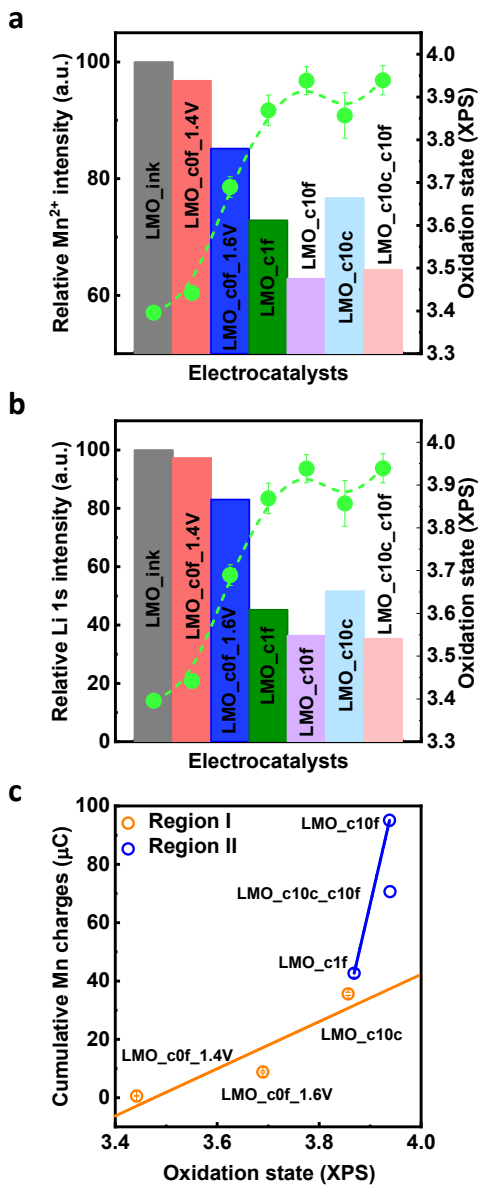

**Figure 5**: (a,b) Normalized relative $Mn^{2+}$ and $Li^+$ intensity of all electrochemically processed samples with respect to LMO_ink and corresponding oxidation state obtained from Mn3s XPS. (c) Correlation of cumulative Mn charges with near-surface oxidation state in Region I and Region II.

**Table 2.** Summary from XPS analysis: the information of relative compositional variation in LMO_ink in term of Mn moieties and delithiation with voltage cycling and the corresponding Mn oxidation obtained through XANES and Mn3s XPS analysis. Compositional changes relative to previous applied key point are indicated, where the sequence in without-conditioning experiment is LMO_ink, LMO_c0f_1.4V, LMO_c0f_1.6V, LMO_c1f and LMO_c10f and with conditioning experiment is LMO_ink, LMO_c10c and LMO_c10c_c10f.

| XPS analysis | $Mn^{2+}$ | $Mn^{3+}$ | $Mn^{4+}$ | $Li^+$ | Mn oxidation state from XANES | Mn oxidation state from Mn3s XPS |
|---|---|---|---|---|---|---|
| LMO_ink | Starting state of material | Starting state of material | Starting state of material | Starting state of material | 3.33 (3) | 3.39 (1) |
| LMO_c0f_1.4V | ↓ | 0 | 0 | 0 | 3.48 (4) | 3.44 (1) |
| LMO_c0f_1.6V | ↓↓ | ↓↓ | ↑↑ | ↓↓ | 3.41 (4) | 3.69 (2) |
| LMO_c1f | ↓↓↓ | ↓↓↓ | ↑↑↑ | ↓↓↓ | 3.60 (5) | 3.87 (3) |
| LMO_c10f | ↓↓ | ↓↓ | ↑↑ | ↓↓ | 3.64 (6) | 3.94 (3) |
| LMO_c10c | ↓↓↓ | ↓↓↓ | ↑↑↑ | ↓↓↓ | - | 3.86 (5) |
| LMO_c10c_c10f | ↓↓↓ | ↓ | ↑ | ↓ | - | 3.94 (3) |

*Down arrows indicate decrease ranging from little decrease ("↓") to strong decrease ("↓↓↓") relative to previous applied key point.

*Up arrows indicate increase ranging from little decrease ("↑") to strong increase ("↑↑↑") relative to previous applied key point.

The Mn oxidation states were also quantitatively obtained by analysis of the difference in the Mn 3s multiplet splitting ($\Delta E_{Mn3s}$) from Mn3s HRXPS (Figure S11; Tables 2 and S7) using a published calibration curve.[40] The multiplet splitting ($\Delta E_{Mn3s}$) decreased with higher potential and cycling, indicating the increase in oxidation state. Noteworthily, the effect of oxidation was more pronounced on the surface so that the Mn oxidation state from Mn3s XPS is larger than the bulk oxidation state from XAS analysis (Table 2; Figure S11c), indicating surface oxidation with anodic potential, which has also been seen previously for other Mn oxides.[41] In particular, the $\Delta E_{Mn3s}$ values of 4.69 eV after 10 cycles including the OER with and without separate conditioning (i.e., samples LMO_c10f and LMO_c10c_c10f) closely matched the $MnO_2$ phase (4.70 eV),[40] suggesting a predominant presence of $Mn^{4+}$ on the surface due to the anodic potentials to study the OER.

Comparison of the qualitative trends from Mn2p, Mn3p and Li1s XPS analysis to the quantitative Mn3s XPS analysis indicated that Mn oxidation was charge balanced by both $Mn^{2+}$ (likely tetrahedral) loss[39] (Figure 5a) and $Li^+$ loss (Figure 5b). The reduction in $Mn^{2+}$ and $Li^+$ appeared to be correlated. Notably, the sample after post-conditioning (LMO_c10c_c10f) retained the same amount of $Mn^{2+}$ and $Li^+$ as compared to the sample without conditioning (LMO_c10f), whereas less Mn loss was detected for the former sample (Figure 3). This inspired us to correlate the electrochemical Mn dissolution charges obtained in the previous section with the electronic properties of the surface.

The cumulative Mn charge detected on the ring of an RRDE depended clearly linearly on the near-surface Mn oxidation state where higher oxidation state led to more Mn loss (Figure 5c). Similar trends were observed as function of $Mn^{2+}$ or $Li^+$ concentration in the samples (Figure S12). Our analysis clearly showed different linear trends in Region I and II, i.e., with and without OER, where more Mn loss was detected per change in oxidation state for concomitant OER. These trends highlight the different natures of the two detected Mn dissolution processes as discussed in detail below.

In this report, we have identified two Mn dissolution processes on the cubic spinel $LiMn_2O_4$. Using ex-situ HRXPS, we found surface oxidation with the formation of mixed $Mn^{3+/4+}$ with relatively increasing $Mn^{4+}$ population with applied potentials and cycling. This aligns with prior research indicating Mn surface oxidation to $Mn^{4+}$ beyond 1.2 V vs RHE, occurring even before the OER onset.[24,41,42] The linear trend of Mn loss with Mn oxidation toward $Mn^{4+}$ strongly suggests that both Process I and Process II proceed via the reaction $MnO_2(s) + 4OH^- \rightarrow MnO_4^- + 2H_2O + 3e^-$ ($E^0 = 0.595$ V vs SHE, 1.362 V vs. RHE at pH 13). While the onset potential for Process I of about 1.4 V aligns with the reversible potential of this reaction and experimental observations from other research,[1,14,15] while the mechanics in Region II, i.e., during OER remain ambiguous.

Region II displayed stronger Mn dissolution with minor surface oxidation changes, suggesting an additional process that makes $Mn^{4+}$. It is plausible that this process is the OER where $Mn^{3+}$ is oxidized to $Mn^{4+}$. The OER mechanism of $LiMn_2O_4$ has been extensively studied by our group before.[23] Note that Mn is believed to oxidize independently of the prototypical OER mechanism so that the argument is valid for any known OER mechanism.[43–46] The postulation that $Mn^{3+}$ to $Mn^{4+}$ oxidation during OER causes Mn dissolution is fully consistent with our observations and also explains the reduced Mn loss for optimized conditioning. The conditioning procedure (sample LMO_c10c) leads to a higher Mn oxidation state as compared to a single half cycle without OER (sample LMO_c0f_1.6V), which means less $Mn^{3+}$ before the onset of the OER that can oxidize to $Mn^{4+}$ and thus less Mn loss for an identical number of OER cycles after conditioning (sample LOM_c10c_c10f). Thus, we hypothesize that the same reaction causes Mn loss in both identified regions where the OER enhances the production of the needed $Mn^{4+}$ precursor state in addition to OER-independent surface oxidation of the Mn oxide.

It is evident from our research that a thorough understanding of the atomic origins of Mn dissolution mechanisms is vital for effectively preventing electrocatalyst degradation. The role of $Mn^{3+}$ and $Mn^{4+}$ for stability has been under debate. Rabe et al. found $Mn^{3+}$-based oxide without OER more stable than $Mn^{4+}$-based oxide with OER,[15] while Morita et al and Ramirez et al found lower stability of $Mn^{3+}$-based oxides during OER.[47,48] Our observations suggest that the increased dissolution with rising $Mn^{4+}$ surface species can be additionally produced by the OER from $Mn^{3+}$. Therefore, more robust Mn-based electrocatalysts need to prevent excessive Mn oxidation before and during the OER as we reported for a Mn-Co oxide.[49]

**CONCLUSIONS**

We discussed pristine cubic spinel $LiMn_2O_4$ particles as an electrocatalytic model for the OER in alkaline media for Mn-based battery materials[50,51] and Mn oxides in general, which all are expected to suffer from Mn loss and in the case of the battery materials also Li or Na loss. We used an RRDE, where manganese loss at the disk was qualitatively probed at the ring to deepen our mechanistic understanding of electrocatalysts degradation. Two distinct onsets of exponential increasing current of Mn dissolution were observed. Process I above 1.4 V with Tafel slopes around 145±15 mV/dec and Process II above 1.6 V vs. RHE with Tafel slopes decreasing from 520±44 mV/dec to 240±18 mV/dec. While Mn dissolution with the former onset is discussed to some detail in the literature,[1,14,15] the distinct Mn dissolution process during the OER is less understood. To close this knowledge gap, we performed XAS and XPS measurements to gain insights into the atomic origins of these electrochemical differences. The data shows no indication of foreign phases or phase changes. The bulk of the samples oxidized from +3.33(3) for the ink-cast electrode to +3.64(6) after 10 cycles, where with application of higher potentials and cycling led to more oxidation. Both the Mn-O as well as Mn-Mn distances also decreased as expected for Mn oxidation. Mn3s XPS analysis revealed a larger extent of oxidation near the surface, ranging from +3.39(1) for the ink-cast electrode to +3.94(3) after 10 cycles. Additionally, the XPS data qualitatively showed that the Mn oxidation with voltage cycling was balanced by both $Mn^{2+}$ and Li[+] ions. This physical characterization enabled us to correlate the cumulate charge associated with Mn dissolution from RRDE experiments to the near-surface oxidation state of samples subjected to various conditions and cycles. Mn loss increased monotonously with oxidation of the Mn remaining in LiMn$_2$O$_4$, where two linear trends were identified. Less Mn was lost per increase in oxidation state without OER as compared to with concomitant OER, suggesting a distinct nature of the two Mn dissolution processes.

The comprehensive understanding of Mn dissolution from the LiMn$_2$O$_4$ model system shows the pivotal role of Mn$^{4+}$ species in the dissolution process. From correlation between surface electronic properties and dissolution, we concluded that increased population of Mn$^{4+}$ species at the surface influences the Mn loss in both identified regions. The increased dissolution trend during OER suggested the formation of additional Mn$^{4+}$ from Mn$^{3+}$ during OER alongside surface oxidation. This postulation suggests that Mn oxides must be made robust against Mn oxidation of two distinct processes with separate onsets. It must be noted that the OER likely requires some Mn$^{4+}$, which would anticorrelate with stability and activity.[47,52] Nonetheless, we performed additional experiments where near-surface Mn oxidation to +3.86(5) was induced without OER to reduce the concentration of Mn$^{3+}$ species that could oxidize during OER. This conditioning treatment resulted in 47% less Mn loss during OER cycling without significant activity loss as compared to direct OER cycling without previous conditioning by Mn oxidation. Based on our observations, we expect that the typical conditioning step of electrocatalysts has a large impact on the degradation as well as potentially the activity of Mn oxides. Our work showed that atomistic insights into the Mn dissolution processes are crucial for knowledge-guided mitigation of electrocatalyst degradation. We expect that the general insight into the role of Mn$^{4+}$ for degradation of Mn oxides can be extended widely to manganese-based oxide systems.

## Materials and Methods

Nano-sized (>99%) $LiMn_2O_4$ were purchased from Sigma-Aldrich. Tetrahydrofuran (THF) was purchased from VWR (≥99.9% stabilized). The electrolyte utilized was 1 M sodium hydroxide Titripur (Merck) solution, which was further diluted to 0.1 M using ultrapure water (Milli-Q with a resistivity of ≥18.2 MΩ). The electrolyte was purged with gases (argon 5.0 and oxygen 4.8) obtained from AirLiquide Alphagaz. Acetylene carbon black was purchased from Alfa Aesar (≥99.9%) and was acid-treated.

## Experimental Methods

### Electrochemical experimental Setup

For all electrocatalytic experiments, we employed an OrigaFlex system, which consisted of three OGF500 Potentiostats (Origalys SAS). Our measurement setup included a RRDE-3A rotator (ALS Japan Co Ltd.) and custom-made electrochemical cells made of polytetrafluoroethylene (PTFE) cylinders. The electrochemical cells were configured in a three-electrode configuration with a saturated calomel electrode (SCE) (ALS Japan Co Ltd., RE-2B) as a reference electrode, and platinum wire as a counter electrode. The RRDE-electrode setup consisted of a 4 mm diameter glassy carbon disk (area= 0.126 $cm^2$) and a concentric platinum ring with inner and outer diameters of 5 mm and 7 mm, respectively. The working electrodes (disk and Pt-ring) were separately polished to a mirror finish using $Al_2O_3$- micro polish and cleaned with water and isopropanol. The SCE reference electrode was calibrated daily against a commercial reversible hydrogen electrode (RHE, HydroFlex Gaskatel).

**Electrochemical experimental Setup**

The recipe of our experiment has been taken from the our previous publication.[25]

To calibrate the ring of the RRDE, a 0.1 M NaOH electrolyte was first saturated with oxygen and a cyclic voltammetry (CV) experiment was conducted at the ring electrode in the potential range of 0-1.75 V vs. RHE at the scan rate of 100 mV/s for five cycles at a rotation speed of 1600 rpm. The same procedure was repeated for an electrolyte saturated with argon. Subsequently, 10.3 mg of $KMnO_4$ was added to the argon-saturated electrolyte for a final molarity of 1 mM and another CV was performed.

For degradation studies, a catalytic ink of $LiMn_2O_4$ was prepared by mixing 10 mg of $LiMn_2O_4$ powder and 2 mg of carbon black into a 2 mL of tetrahydrofuran (THF) slurry. After 30 minutes of sonication, 10 μL of this suspension was applied on a polished glassy carbon disk, which was then assembled in a RRDE setup. The ink was uniformly spread across the entire disk surface. The electrolyte was also saturated with argon 30 minutes prior to any of the electrochemical measurements in Ar-saturated 0.1 M NaOH

**Electrochemical measurements**

Three measurements were carried out by preparing the $LiMn_2O_4$ nanoparticles drop-casted over glassy carbon disk electrodes and assembled the RRDE setup. The total oxidation current was recorded on disk, whereas oxygen and manganese were detected on the ring. The ring currents shown here are the uncalibrated current. In our experiments, the disk electrode was conditioned at 1.25 V vs. RHE for five minutes and then CV were performed in the potential window of 1.25 to 1.75 V vs RHE. All measurements were carried out with the scan rate of 10 mV/s and with the rotation speed of 1600 rpm. The oxygen and manganese detection current has been measured by applying the 0.4 V and 1.2 V vs RHE, respectively. All characterization and electrochemical experiments were carried out at room temperature. Subsequently, the electrochemical impedance spectroscopy (EIS) was employed in the frequency range of 100 kHz to 1 Hz to determine the uncompensated resistance and correct the applied voltage for the ohmic drop. The electrode coated with nano sized $LiMn_2O_4$ exhibited a typical uncompensated resistance.

**Physical characterization of LiMn₂O₄ particles**

**X-ray diffraction (XRD)**

We examined the crystal structure of pristine $LiMn_2O_4$ particles using θ-2θ X-ray diffraction (XRD) on a Bruker D8 diffractometer equipped with a Cu Kα source. We did not study the electrochemically processed samples using XRD because the available material was insufficient for a clear signal.

**X-ray photoelectron spectroscopy (XPS)**

We investigated near-surface elements and local bonding using XPS measurements on a Kratos Axis Supra instrument, which is equipped with a monochromated Al Kα X-ray source. Wide XPS spectra were performed on all samples with a pass energy of 160 eV and a binding energy step size of 1 eV. We carried out high-resolution (HRXPS) scans on the Mn 2p, C 1s, Mn 3p/Li 1s and Mn 3s edges, using a step size of 0.05 eV and a pass energy of 5 eV. The multiplet splitting of Mn 3s HRXPS peaks ($\Delta E_{Mn3s}$) were used to determine the oxidation state of Mn. The calibration curve was derived from the $\Delta E_{Mn3s}$ values of Mn oxides with known oxidation state.[53]

**X-ray absorption spectroscopy (XAS)**

All XAS data was collected using an averaged nominal ring current of 300 mA in both transmission and fluorescence modes at the BESSY II synchrotron, which is operated by Helmholtz-Zentrum Berlin.

At KMC-2, the general used setup was arranged as follows: $I_0$ ionization chamber, sample, $I_1$ ionization chamber or FY detector, energy reference and $I_2$ ionization chamber. The used double monochromator consisted of two Ge-graded Si(111) crystal substrates and the polarization of the beam was linear horizontal.[54] To prepare the reference samples, a thin and uniform layer of the powder was spread on Kapton tape. After removing the surplus powder, the tape was folded multiple times to create 1 cm x 1 cm windows. These reference samples ($MnO$, $Mn_3O_4$, $Mn_2O_3$, $LiMnO_2$, $LiMn_2O_4$ and $MnO_2$) were evaluated in transmission mode between two ion chambers detectors at ambient temperature. Electrochemically processed samples were prepared by dropcasting the $LiMn_2O_4$ ink on 4mm glassy carbon disk and evaluated in fluorescence mode. The reference Mn-foil is also simultaneously collected corresponding to each measurement. Two refences spectra were compared to confirm the correct energy calibration. For the X-ray absorption near edge structure (XANES) energy calibration, the respective metal foil was utilized, setting the inflection point for Mn at 6539 eV. All spectra normalization involved the subtraction of a straight line obtained by fitting the data before the K-edge and division by a polynomial function obtained by fitting the data after the K-edge.

The Fourier transform (FT) of the extended X-ray absorption fine structure (EXAFS) was determined between 44 and 412.1 eV (3.4 to 10.4 A) above the K edge, with $E_0$ values set at 6539 eV for Mn. EXAFS simulation gives the structural information of three of relevant parameters; co-ordination number (N) which is related to the number of neighboring atoms around the absorber atom, averaged interatomic distance (R) is nothing but the bond distance between the absorber atom and scatterer and Debye-Waller factor ($\sigma$) which is associated with the distance distribution in a disordered material. It must be noted that, Mn-O and Mn-Mn co-ordination peak position on the reduced distance scale are approximately 0.3 Å smaller than the true interatomic distances. We performed EXAFS simulations using the SimXLite software. The phase functions were calculated with the FEFF8-Lite program (version 8.5.3) with the self-consistent field option enabled. The used phase functions were simulated using the $LiMn_2O_4$ crystal structure from the crystallography open database with ID 1513966.[27] We fixed N=6 for both Mn-Mn and Mn-O coordination path and fit all other parameters including amplitude reduction factor ($S0^2$). To optimize the EXAFS simulations, we minimized the error sum which was obtained by summing the squared deviations between the measured and simulated values, i.e., least-squares fit approach. The R-factor was estimated by Fourier filtering data and fit model between of 0 and 3 Å. The fitting was conducted using the Levenberg-Marquardt method with numerical derivatives.[54]

AUTHOR INFORMATION

**Corresponding Author**

*marcel.risch@helmholtz-berlin.de

**Funding Sources**

Deutsche Forschungsgemeinschaft (DFG) - Projektnummer 217133147.

**Data availability**
The data that support the findings of this study are openly available in Zenodo at https://doi.org/10.5281/zenodo.8306360, reference number 8306360.

**Conflicts of interest**

There are no conflicts to declare.

ACKNOWLEDGMENT

This work was financially supported by SFB 1073 (project C05) funded by the Deutsche Forschungsgemeinschaft (DFG, German Research Foundation) – project number 217133147. We thank the Helmholtz-Zentrum Berlin für Materialien und Energie for the allocation of synchrotron radiation beamtime. The authors thank to Joaquín Morales-Santelices, Denis Antipin, Julian Lorenz, Corinna Harms, Konstantin Rücker and Dereje Hailu Taffa for helping with XAS data collection, Frederik Stender for remote help during beamline and Götz Schuck for support at the KMC-2 beamline. The authors also thank Florian Schönewald for helping with XPS data collection. The XPS instrument was funded by the DFG.

# Supporting Information

# Manganese Dissolution in alkaline medium with and without concurrent oxygen evolution in LiMn$_2$O$_4$

*Omeshwari Bisen[¶†], Max Baumung[¶†], Cynthia A. Volkert[†], Marcel Risch[¶†]\**

† Institute of Materials Physics, University of Göttingen, Friedrich-Hund-Platz 1, 37077 Göttingen, Germany.

¶ Nachwuchsgruppe Gestaltung des Sauerstoffentwicklungsmechanismus, Helmholtz Zentrum Berlin für Materialien und Energie GmbH, Hahn-Meitner-Platz 1, 14109 Berlin, Germany.

\* corresponding author: marcel.risch@helmholtz-berlin.de

This supporting information includes 12 figures and 7 tables.

**Supporting Figures**

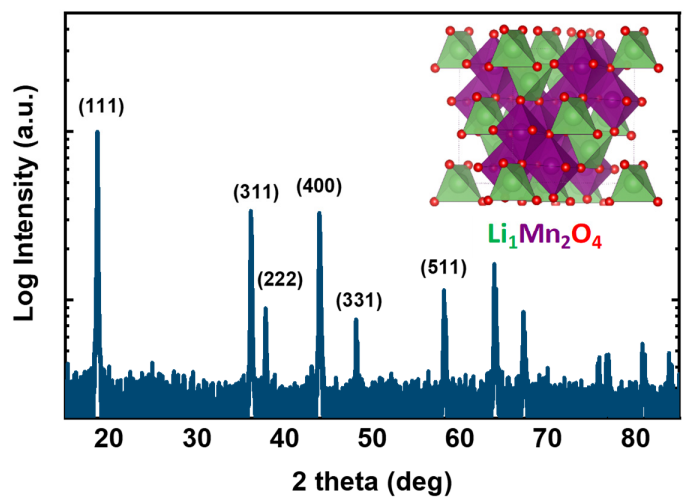

**Figure S1**. Indexed X-ray diffraction pattern of pristine LiMn$_2$O$_4$ particles (logarithmic scale). The inset shows a cubic spinel crystal structure of LiMn$_2$O$_4$.

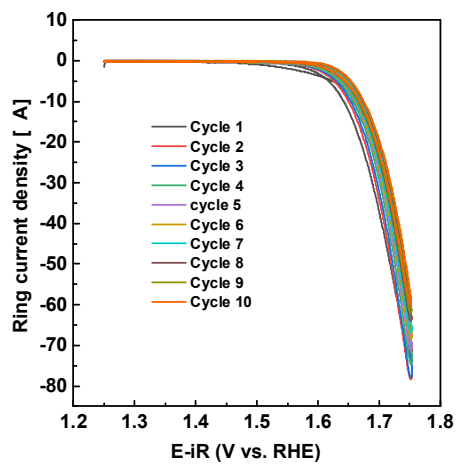

**Figure S2.** As measured oxygen evolution current observed at Pt ring (potential applied at ring is 0.4 V vs RHE) for 1$^{st}$ to 10$^{th}$ cycle.

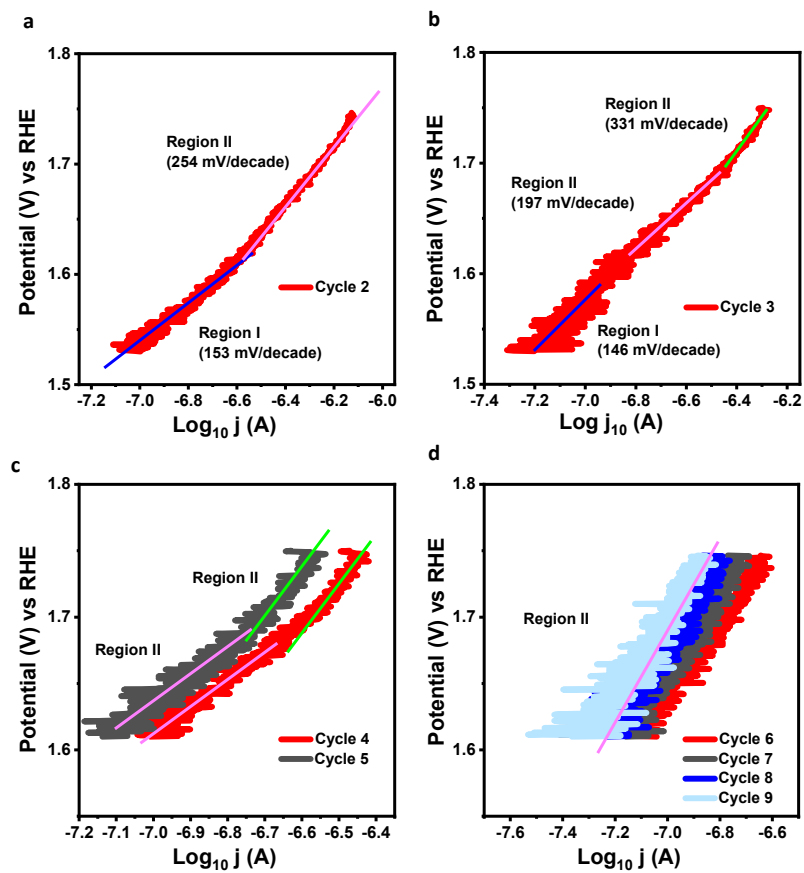

**Figure S3.** (a-d) Tafel slope of 2[nd] to 9[th] anodic cycle of Mn dissolution current observed at Pt ring. Tafel slopes obtained in region I (1.4 < E < 1.6 V vs RHE) and region II (1.6 < E < 1.75 V vs RHE).

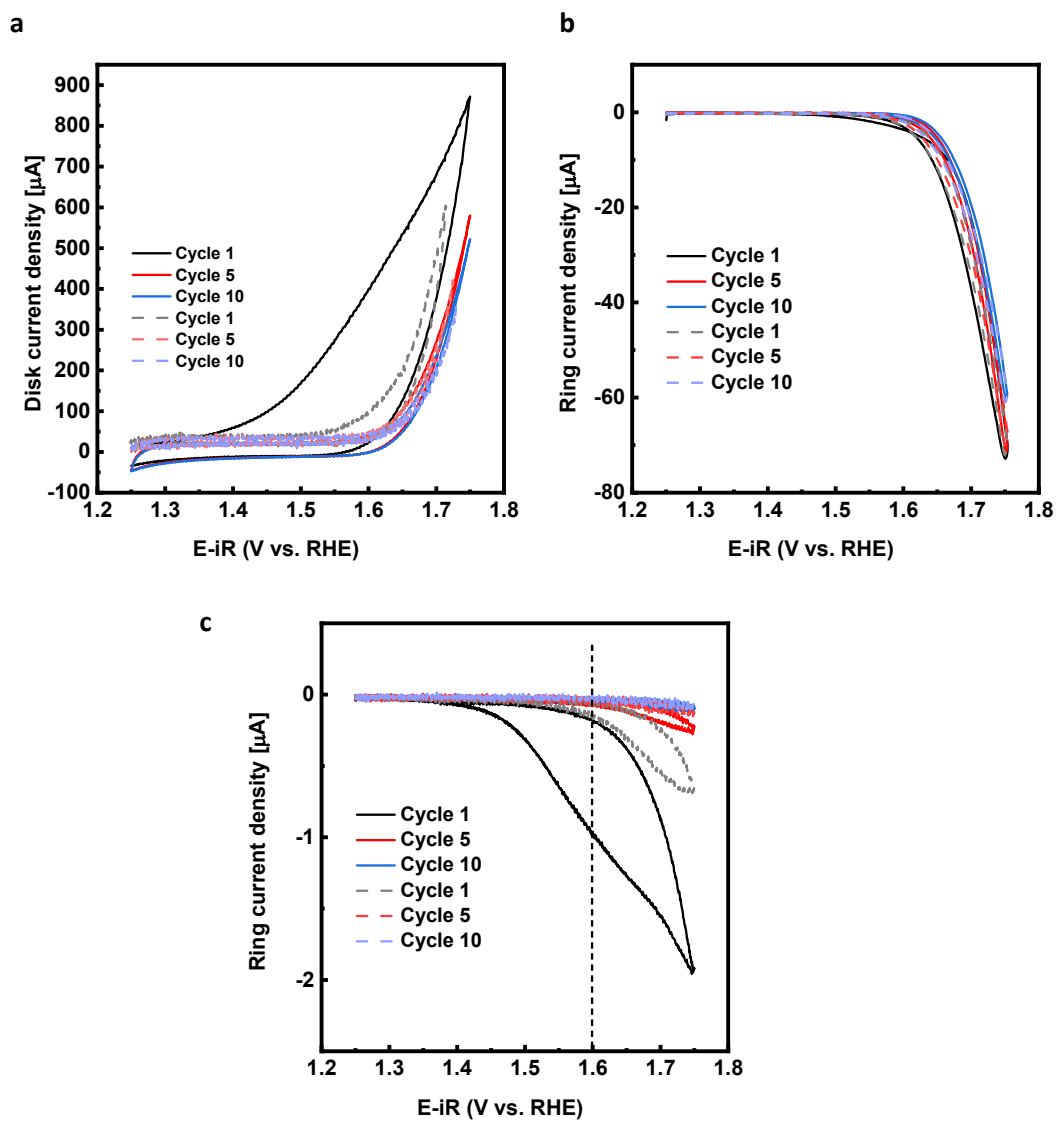

**Figure S4**. (a-c) CV at disk for cycle 1, 5 and 10. (b,c) CV at ring for $O_2$ and Mn detection for cycle 1, 5, 10, respectively. Dotted and solid lines showed the CVs with and without conditioning.

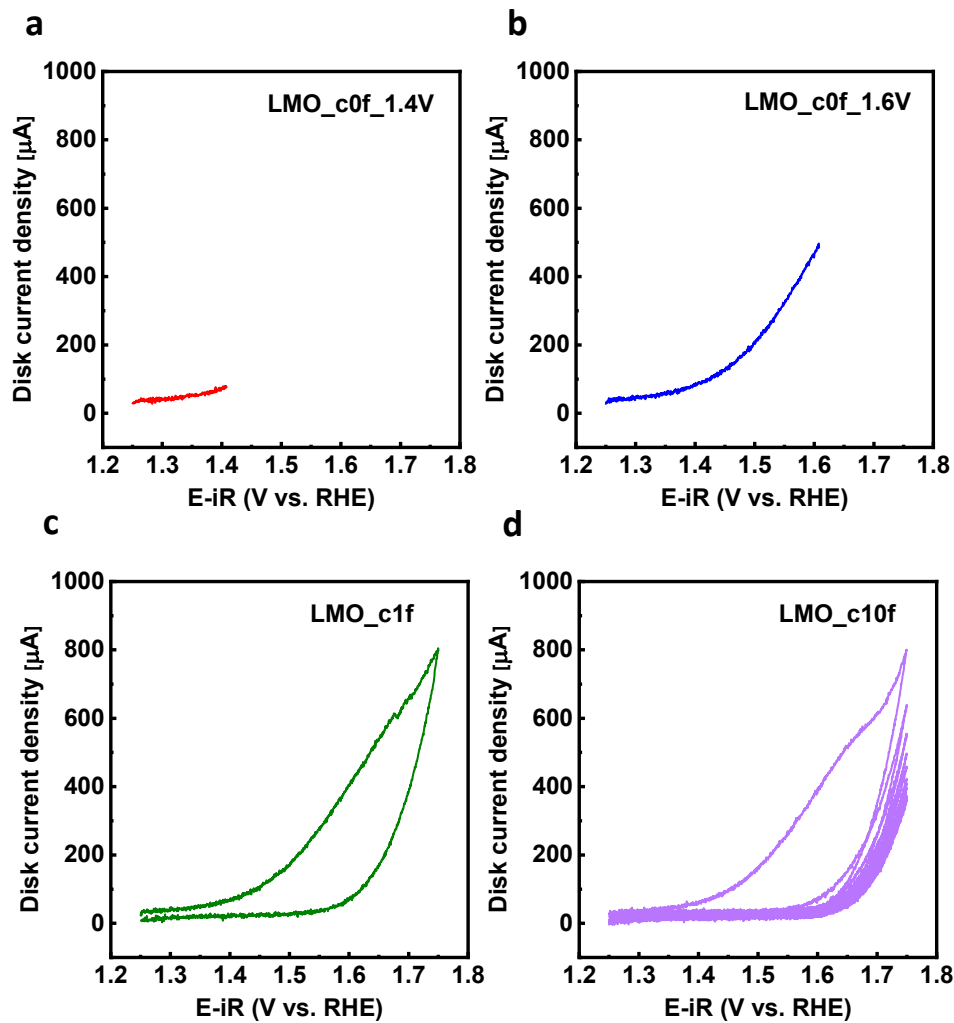

**Figure S5.** Electrochemical measurements for systematic XAS and XPS analysis. (1) LMO_c0f_1.4V: drop-casted particles scan up to the potential of 1.4 V vs RHE, (2) LMO_c0f_1.6 V: scans up to 1.6 V vs RHE, (3) LMO_c1f: after completion of 1 potential cycle in the potential range of 1.25-1.75 V vs RHE, (4) LMO_c10f: after 10 potential cycles in the potential range of 1.25-1.75 V vs RHE.

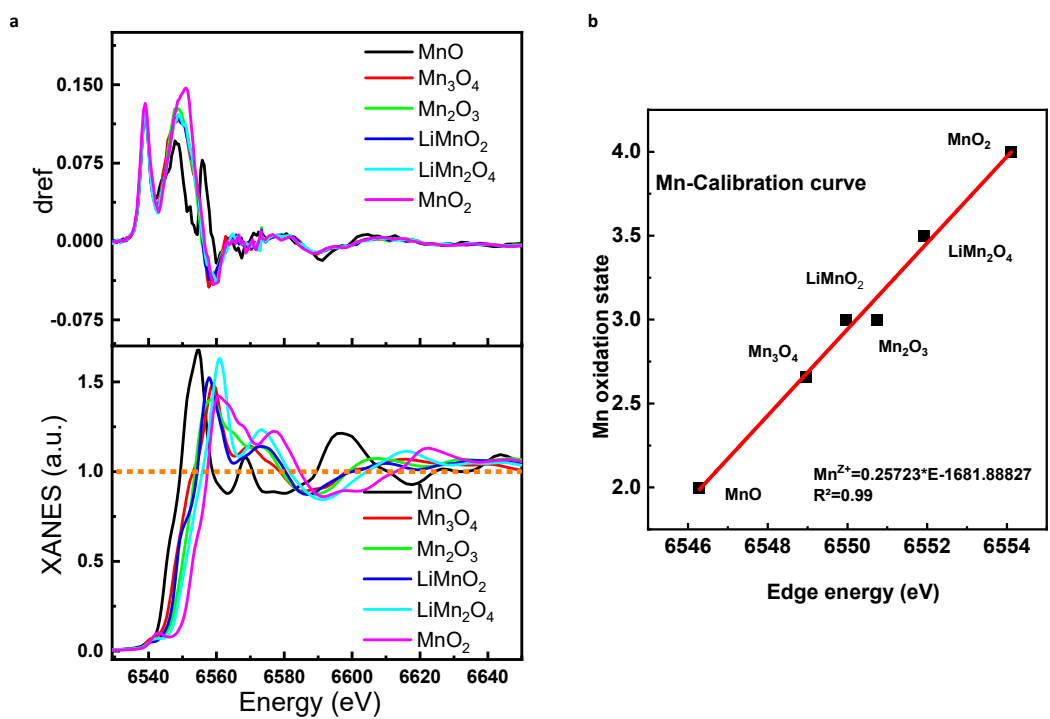

**Figure S6.** Mn-K edge XANES spectra and derivative of Mn reference foil collected during each XAS measurements of Mn-based reference; MnO, $Mn_3O_4$, $Mn_2O_3$, $LiMnO_2$, $LiMn_2O_4$, $MnO_2$. (b) Mn calibration curve obtained from nominal oxidation state of Mn-references as a function of energy of the Mn-K edge. The estimated oxidation states are shown in Table S2. The edge energy was estimated using the integral method ($\mu_1$=1.00, $\mu_2$=0.20).

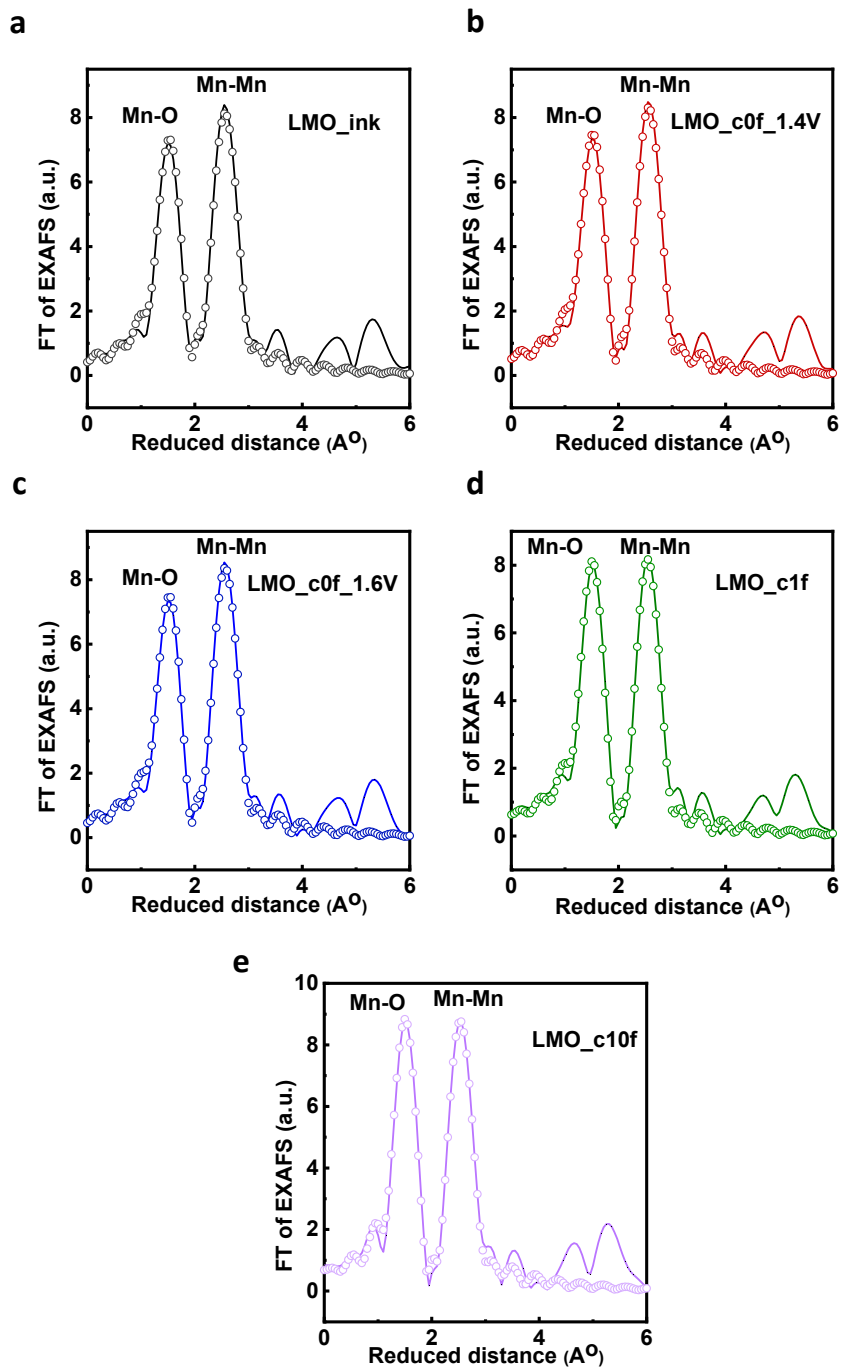

**Figure S7.** (a-e) FT of EXAFS spectra of LMO_ink and electrochemically processed sample LMO_c0f_1.4V, LMO_c0f_1.6V, LMO_c1f and LMO_c10f at Mn-K edge, where the solid lines represent the measurements and scattered dotted lines are respective EXAFS simulations.

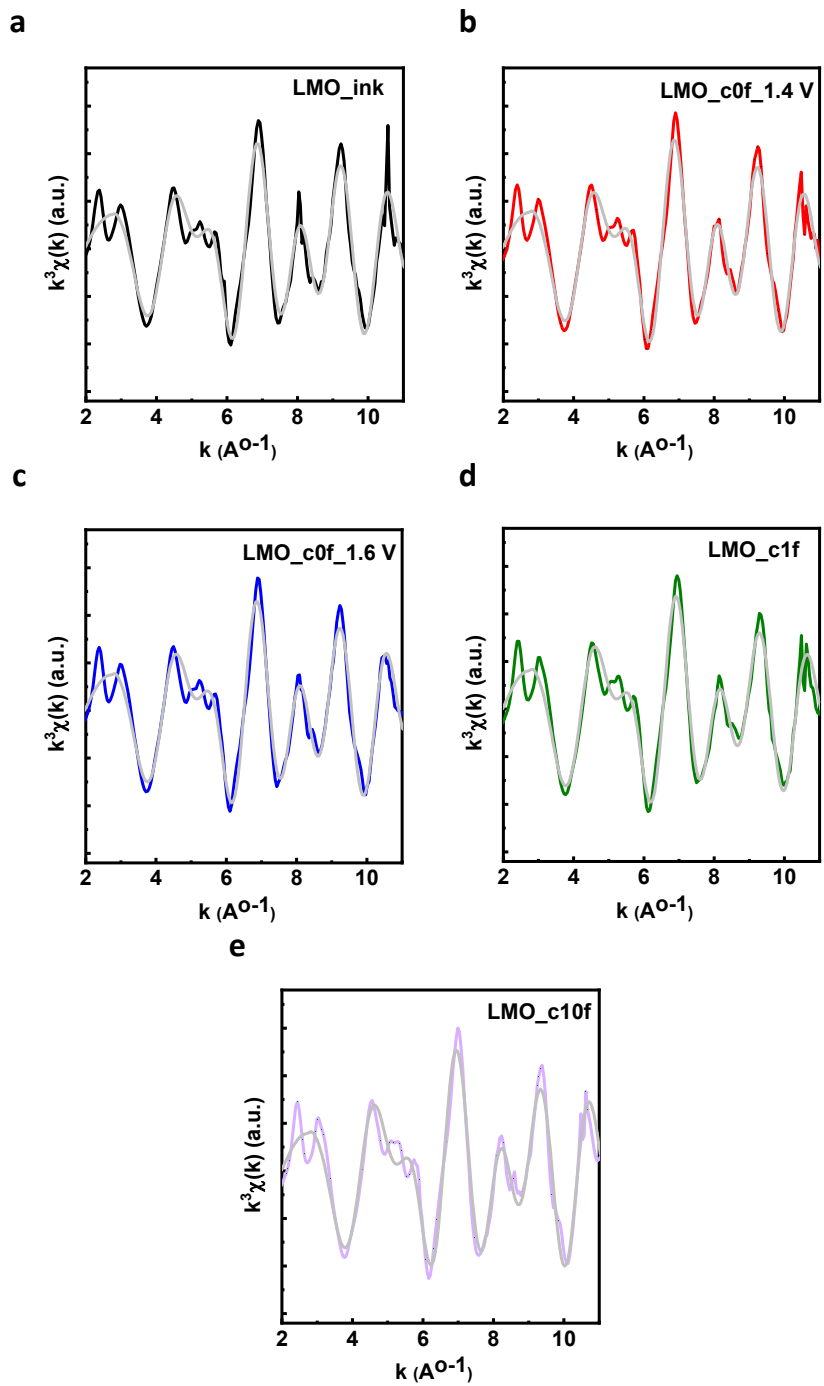

**Figure S8.** $k^3$-weighted EXAFS spectra of LMO_ink and electrochemically processed samples LMO_c0f_1.4V, LMO_c0f_1.6V, LMO_c1f and LMO_c10f recorded at the Mn-K edge. The colorful lines represent the measurements and gray lines are respective EXAFS simulations.

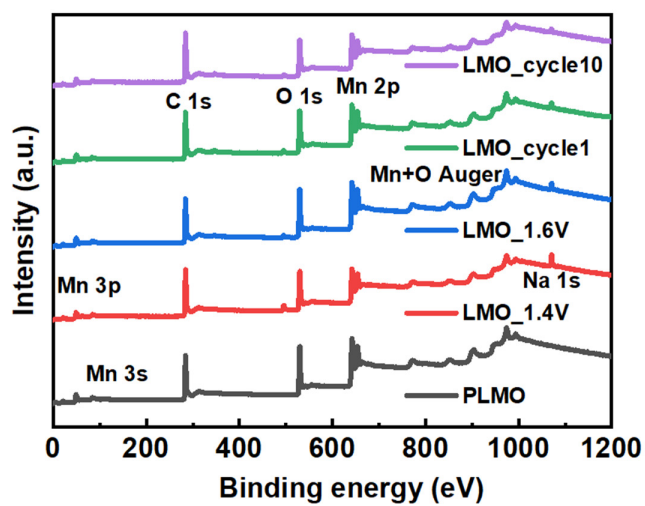

**Figure S9.** Wide XPS spectra of LMO_ink and electrochemically processed samples. All samples were dropcasted on glassy carbon disk.

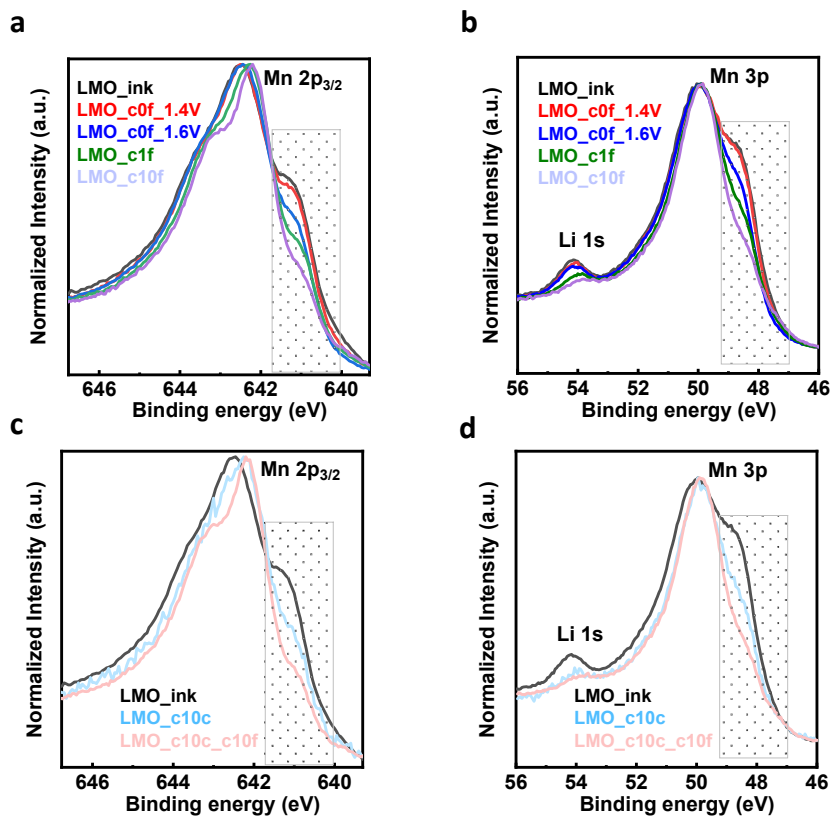

**Figure S10.** (a,b) Mn 2p, Mn 3p and Li 1s (shoulder at higher B.E.) HRXPS of LMO_ink, LMO_c0f_1.4V, LMO_c0f_1.6V, LMO_c1f and LMO_c10f. (c,d) Mn 2p, Mn 3p and Li 1s (shoulder at higher B.E.) HRXPS of LMO_ink, LMO_c10c and LMO_c10c_c10f.

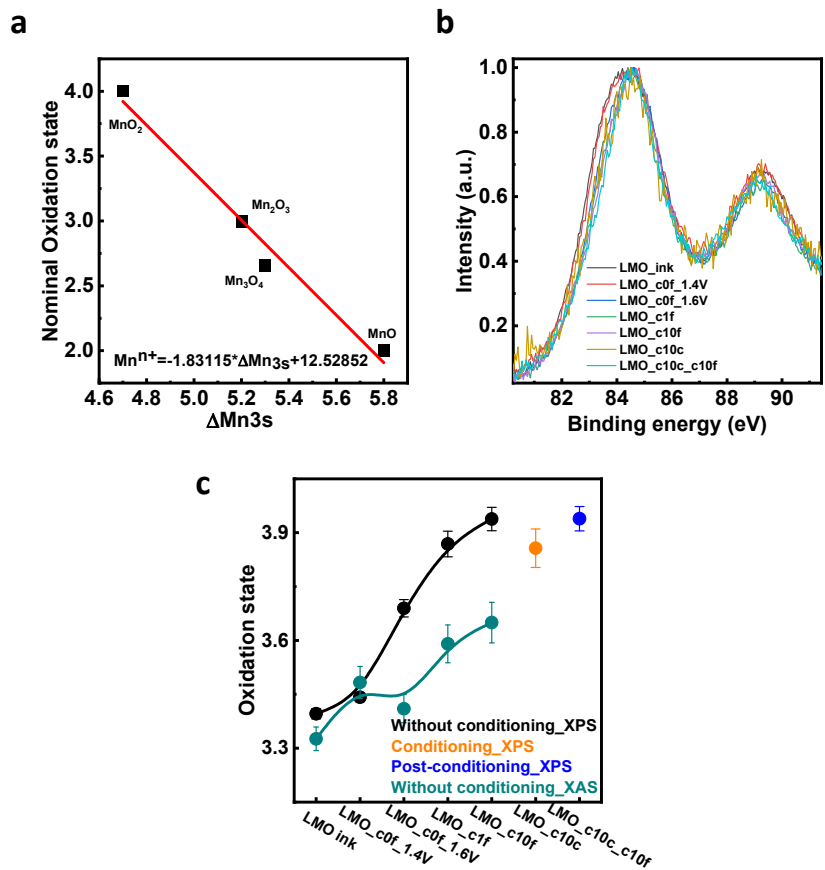

**Figure S11.** (a) Mn calibration curve obtained from nominal oxidation state of Mn-references (MnO (+2), $Mn_3O_4$ (+2.66), $Mn_2O_3$ (+3), $MnO_2$ (+4)) as a function of difference in Mn 3s multiplet splitting. (b) Mn 3s HRXPS spectra of LMO_ink and electrochemically processed samples LMO_c0f_1.4V, LMO_c0f_1.6V, LMO_c1f, LMO_c10f, LMO_c10c and LMO_c10c_c10f. (c) Bulk average oxidation state obtained from Mn-K edge XANES spectra and surface oxidation state from Mn 3s HRXPS spectra for all aforementioned samples.

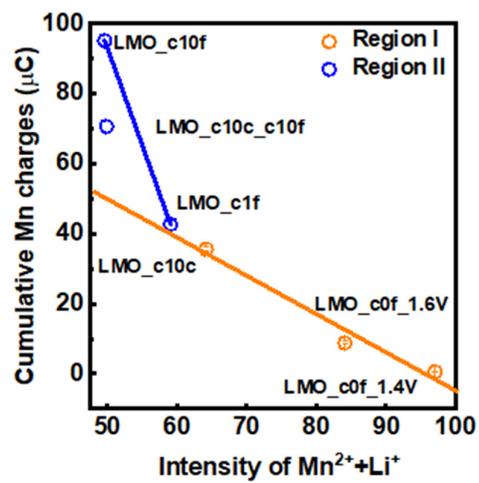

**Figure S12.** Correlation of cumulative Mn charges with intensity of ($Mn^{2+}+Li^+$) in Region I and Region II.

**Supporting Tables**

**Table S1.** Averaged Tafel slope values for 1st to 10th anodic cycle of Mn dissolution current of LMO_ink observed at Pt ring. The values are the average of three different samples and standard deviation is reported as error.

| Cycle | Region I (1.4<E<1.6) | | Region II (1.6<E<1.75 V vs RHE) | | Region II (1.6<E<1.7) | | Region II (1.7<E<1.75) | |
|---|---|---|---|---|---|---|---|---|
| | Average (mV/dec) | Error (mV/dec) | Average (mV/dec) | Error (mV/dec) | Average (mV/dec) | Error (mV/dec) | Average (mV/dec) | Error (mV/dec) |
| 1 | 139 | 4 | 520 | 44 | | | | |
| 2 | 149 | 5 | 254 | 10 | | | | |
| 3 | 146 | 13 | 229 | 3 | 198 | 2 | 331 | 21 |
| 4 | | | 223 | 1 | 193 | 2 | 292 | 33 |
| 5 | | | 231 | 4 | 194 | 8 | 255 | 16 |
| 6 | | | 239 | 6 | | | | |
| 7 | | | 237 | 8 | | | | |
| 8 | | | 253 | 8 | | | | |
| 9 | | | 246 | 7 | | | | |
| 10 | | | 243 | 1 | | | | |

**Table S2.** Averaged Tafel slope values for 1st to 10th anodic cycle of Mn dissolution current of LMO_ink during conditioning step (in the potential window of 1.25-1.6 V vs RHE). Tafel slope were plotted in region I (1.4 < E <1.6 V vs RHE). The values are the average of three different samples and standard deviation is reported as error.

| Cycle | Region I (1.4<E<1.6) | |
|-------|-----------------------|-----------------|
|       | Average (mV/dec)      | Error (mV/dec)  |
| 1     | 144                   | 8               |
| 2     | 129                   | 0.5             |
| 3     | 130                   | 4               |
| 4     | 114                   | 14              |
| 5     | 120                   | 3               |
| 6     | 112                   | 5               |
| 7     | 122                   | 3               |
| 8     | 115                   | 7               |
| 9     | 102                   | 9               |
| 10    | 98                    | 0.4             |

**Table S3.** Averaged Tafel slope values for $1^{st}$ to $10^{th}$ anodic cycle of Mn dissolution current of LMO_ink after conditioning step observed at Pt ring (in the potential window of 1.6-1.75 V vs RHE). The values are the average of three different samples and standard deviation is reported as error.

| Cycle | Region II (1.6<E<1.75 V vs RHE) | | Region II (1.6<E<1.7) | | Region II (1.7<E<1.75) | |
|---|---|---|---|---|---|---|
| | Average (mV/dec) | Error (mV/dec) | Average (mV/dec) | Error (mV/dec) | Average (mV/dec) | Error (mV/dec) |
| 1 | 194 | 16 | 165 | 13 | 328 | 26 |
| 2 | 205 | 16 | 171 | 16 | 257 | 20 |
| 3 | 219 | 16 | 171 | 11 | 215 | 25 |
| 4 | 209 | 20 | 162 | 16 | 202 | 16 |
| 5 | 214 | 27 | | | | |
| 6 | 196 | 21 | | | | |
| 7 | 189 | 20 | | | | |
| 8 | 196 | 2 | | | | |
| 9 | 167 | 17 | | | | |
| 10 | 173 | 1 | | | | |

**Table S4.** The Mn charge passed corresponding to Mn ring current in 1.25-1.75 V vs RHE with and without conditioning step, where conditioning step is the 10 CV in the potential window of 1.25-1.6 V vs RHE. The Mn charge has been evaluated from Figure 1b (without-conditioning), Figure 2a (during conditioning step) and Figure 2c (post-conditioning).

| Cycles | Mn charge (µF) in 1.25-1.75 V (without-conditioning) | | Mn charge (µF) in 1.25-1.75 V (with activation step) | |
|---|---|---|---|---|
| | Mn charge in 1.25-1.6 V (without-conditioning) | Mn charge in 1.25-1.75 V (without-conditioning) | Mn charge in 1.25-1.6 V (during conditioning) | Mn charge in 1.25-1.75 V (post-conditioning) |
| 1 | 1.13E-05 | 4.27E-05 | 1.13E-05 | 1.16E-05 |
| 2 | 3.78E-06 | 1.53E-05 | 5.79E-06 | 5.17E-06 |
| 3 | 2.06E-06 | 8.48E-06 | 3.83E-06 | 3.55E-06 |
| 4 | 1.64E-06 | 5.96E-06 | 3.01E-06 | 2.84E-06 |
| 5 | 1.18E-06 | 4.50E-06 | 2.49E-06 | 2.51E-06 |
| 6 | 1.17E-06 | 3.75E-06 | 2.23E-06 | 2.33E-06 |
| 7 | 9.43E-07 | 3.05E-06 | 2.05E-06 | 2.08E-06 |
| 8 | 9.11E-07 | 2.76E-06 | 1.76E-06 | 1.95E-06 |
| 9 | 1.10E-06 | 2.59E-06 | 1.66E-06 | 1.73E-06 |
| 10 | 1.03E-06 | 2.42E-06 | 1.51E-06 | 1.34E-06 |
| Total charge | **2.51E-05** | **9.15E-5** | **3.56E-5** | **3.50E-5** |
| | | | 3.44E-5 + 3.75E-5 = **7.06E-5** | |

**Without conditioning**

Charge in Region I=2.51 e-05 µF, Charge in Region II=6.64 E-05 µF

**With conditioning**

Charge in Region I=3.56 e-05 µF, Charge in Region II=3.50 E-05 µF

% change in charge in Region II= ((6.64 E-05 µF−3.50 E-05 µF) / 6.64 E-05 µF) *100 = ~ 47.29%

% change in total charge = ((9.15 E-05 µF−7.06 E-05 µF) / 9.15 E-05 µF) *100 = ~ 23%

**Table S5.** Edge energy and Mn nominal oxidation state obtained from Mn K-edge XANES spectra of LMO_ink and electrochemically processed samples. The fit equation and graph are shown in Figure S6.

| Sample | Edge energy (eV) | Oxidation state |
|---|---|---|
| LMO_ink | 6551.4 | 3.33 (3) |
| LMO_c0f_1.4V | 6551.99 | 3.48 (4) |
| LMO_c0f_1.6V | 6551.71 | 3.41 (4) |
| LMO_c1f | 6552.47 | 3.60 (5) |
| LMO_c10f | 6552.60 | 3.64 (6) |

**Table S6:** EXAFS absorber-scatter averaged interatomic distance (R), co-ordination number

| Sample | Parameter | Mn-O | Mn-Mn | S0$^2$ | R-factor |
|---|---|---|---|---|---|
| **LMO_ink** | *N* | *6* | *6* | 0.533 | 0.80% |
| | R (Å) | 1.899 | 2.898 | | |
| | σ (Å) | 0.0574 | 0.0589 | | |
| **LMO_c0f_1.4V** | *N* | *6* | *6* | 0.575 | 0.54% |
| | R (Å) | 1.898 | 2.896 | | |
| | σ (Å) | 0.063 | 0.0627 | | |
| **LMO_c0f_1.6V** | *N* | *6* | *6* | 0.566 | 0.54% |
| | R (Å) | 1.899 | 2.896 | | |
| | σ (Å) | 0.0614 | 0.061 | | |
| **LMO_c1f** | *N* | *6* | *6* | 0.597 | 0.52% |
| | R (Å) | 1.893 | 2.876 | | |
| | σ (Å) | 0.0597 | 0.0683 | | |
| **LMO_c10f** | *N* | *6* | *6* | 0.575 | 0.60% |
| | R (Å) | 1.890 | 2.861 | | |
| | σ (Å) | 0.0488 | 0.0625 | | |

The Debye-Waller factor (σ) as determined by simulation of the k$^3$-weighted EXAFS spectra at the Mn K edge for LMO_ink and electrochemically processed samples LMO_c0f_1.4V, LMO_c0f_1.6V, LMO_c1f and LMO_c10f. R-factor and amplitude reduction factor (S0$^2$) are also listed in table.

The number of interactions N was fixed to 6.

**Table S7.** $\Delta E_{Mn3s}$ and Mn surface oxidation state obtained from Mn 3s HRXPS spectra of LMO_ink and electrochemically processed samples. The fit equation and graph are shown in Figure S11.

| Sample | $\Delta E_{Mn3s}$ | Oxidation state |
|---|---|---|
| LMO_ink | 4.99 | 3.39 (1) |
| LMO_c0f_1.4V | 4.96 | 3.44 (1) |
| LMO_c0f_1.6V | 4.83 | 3.69 (2) |
| LMO_c1f | 4.73 | 3.87 (3) |
| LMO_c10f | 4.69 | 3.94 (3) |
| LMO_c10c | 4.74 | 3.86 (5) |
| LMO_c10c_c10f | 4.69 | 3.94 (3) |